\documentclass[fleqn,10pt]{wlscirep}
\usepackage[utf8]{inputenc}
\usepackage[T1]{fontenc}
\usepackage{cite}
\usepackage{amsmath,amssymb,amsfonts}
\usepackage{algorithmic}
\usepackage{graphicx}
\usepackage{textcomp}
\usepackage{multirow}
\usepackage{float}
\usepackage{dblfloatfix}   
\usepackage{kantlipsum}   
\usepackage{hyperref}
\usepackage{siunitx}
\usepackage{longtable}
\usepackage{array}
\title{A novel open-source ultrasound dataset with deep learning benchmarks for spinal cord injury localization and anatomical segmentation}

\author[1]{Avisha Kumar}
\author[1]{Kunal Kotkar}
\author[2,+]{Kelly Jiang}
\author[2,+]{Meghana Bhimreddy}
\author[2,+]{Daniel Davidar}
\author[2,+]{Carly Weber-Levine}
\author[2]{Siddharth Krishnan}
\author[2]{Max J. Kerensky}
\author[1]{Ruixing Liang}
\author[2]{Kelley K. Leadingham}
\author[2]{Denis Routkevitch}
\author[2]{Andrew M. Hersh}
\author[2]{Kimberly Ashayeri}
\author[2]{Betty Tyler}
\author[2]{Ian Suk}
\author[3]{Jennifer Son}
\author[2]{Nicholas Theodore}
\author[1, 2]{Nitish Thakor}
\author[1, 2*]{Amir Manbachi}

\affil[1]{Johns Hopkins University, Baltimore, MD, USA}
\affil[2]{Johns Hopkins University School of Medicine, Baltimore, MD, USA}
\affil[3]{Cleveland Clinic, Cleveland, OH, USA}
\affil[*]{akumar80@jhu.edu}
\affil[+]{these authors contributed equally to this work}


\begin{abstract}
While deep learning has catalyzed breakthroughs across numerous domains, its broader adoption in clinical settings is inhibited by the costly and time-intensive nature of data acquisition and annotation. To further facilitate medical machine learning, we present an ultrasound dataset of 10,223 Brightness-mode (B-mode) images consisting of sagittal slices of porcine spinal cords (N=25) before and after a contusion injury. We additionally benchmark the performance metrics of several state-of-the-art object detection algorithms to localize the site of injury and semantic segmentation models to label the anatomy for comparison and creation of task-specific architectures. Finally, we evaluate the zero-shot generalization capabilities of the segmentation models on human ultrasound spinal cord images to determine whether training on our porcine dataset is sufficient for accurately interpreting human data. Our results show that the YOLOv8 detection model outperforms all evaluated models for injury localization, achieving a mean Average Precision (mAP50-95) score of 0.606. Segmentation metrics indicate that the DeepLabv3 segmentation model achieves the highest accuracy on unseen porcine anatomy, with a Mean Dice score of 0.587, while SAMed achieves the highest Mean Dice score generalizing to human anatomy (0.445). To the best of our knowledge, this is the largest annotated dataset of spinal cord ultrasound images made publicly available to researchers and medical professionals, as well as the first public report of object detection and segmentation architectures to assess anatomical markers in the spinal cord for methodology development and clinical applications. 
\end{abstract}
\begin{document}

\flushbottom
\maketitle

\thispagestyle{empty}

\section*{Introduction}
Spinal cord injury (SCI) is a devastating condition affecting an estimated 900,000 people worldwide in 2019 alone, predominantly stemming from motor vehicle accidents and violence \cite{liu2023spinal, ding2022spinal}. The aftermath of SCI extends beyond the acute trauma, as patients experience mobility impairments, severe pain, autonomic dysregulation, and increased risk of mortality if not treated effectively and efficiently \cite{ahuja2017traumatic}. SCI can be divided into 2 phases: the primary injury that results immediately from the traumatic event, and the secondary injury, in which the clinical manifestations of SCI are exacerbated by inflammation and edema, compromising vascular perfusion within the cord \cite{hwang2021ultrasound, tsehay2022advances}. The current treatment approach involves a surgical decompression procedure, during which the vertebral structures surrounding the injury site are removed to promote blood flow \cite{manbachi2020intraoperative}. 
\par
Interventions such as cerebrospinal fluid (CSF) drainage, augmentation of mean arterial pressure goals, and electrical stimulation show promise to mitigate the adverse physiological sequelae following SCI and may improve functional outcome \cite{martirosyan2015cerebrospinal, horn2008effects, schuhmann2019electrical, whiting2018posterior}. However, optimally titrating these therapies remains a challenge due to the lack of real-time and automatic monitoring of spinal cord parameters such as hematoma development and tissue inflammation. Ultrasound is a promising diagnostic tool for SCI management, providing clinicians with real-time, remote, and portable imaging capabilities for visualizing anatomical boundaries and identifying pathological abnormalities within the tissue \cite{lee2017ultrasonic}. Brightness-mode (B-mode) ultrasound emits acoustic waves at frequencies $\geq$ 20 kHz to construct a two-dimensional grayscale image representing the anatomy within the imaging plane \cite{conlon2019moving}. Without the acoustic barrier of the vertebral bone after surgical decompression (e.g., laminectomy), clinicians can perform intraoperative imaging of the spinal cord with ultrasound. This information allows surgeons to verify sufficient decompression of the cord and evaluate the need for additional interventions \cite{aarabi2022proposal}.  
\par
Compared to other widely accepted imaging modalities, like magnetic resonance imaging (MRI), computed tomography (CT), and conventional radiography (X-ray), ultrasound has benefits for rapid detection of tissue irregularities and injury management due to its safety, cost-effectiveness, and real-time imaging \cite{kumar2023visualizing}. The severity and location of SCI can be understood by studying the inflammation in the tissue and the development of a hematoma (i.e., pooling of blood in a tissue due to broken blood vessels from the injury). Detailed knowledge of the injury and its evolution can provide valuable insights on treatment efficacy and patient prognosis, as spinal cord compression, swelling, and hemorrhage are associated with poor prognosis of neurologic recovery \cite{miyanji2007acute, al2011clinical}. Furthermore, monitoring the swelling of the key anatomy surrounding the injury provides new avenues for researching the underlying physiological mechanisms of secondary injury. While ultrasound's real-time capabilities make it a valuable tool for continuous tissue monitoring, its utility is often compromised by poor image quality, signal reverberation, artifacts, and speckle. These imaging challenges require the expertise of skilled sonographers or radiologists to ensure accurate interpretation \cite{ostras2021diagnostic, long2020coherence}. Unfortunately, these image quality issues are a significant contributor to inter-observer variability, which further complicates medical image evaluations. As a result, these factors are key drivers in diagnostic errors underscoring the critical need for medical image algorithms for improving diagnostic results \cite{itri2018fundamentals, pinto2010spectrum}. 
\par 
To take full advantage of continuous imaging with ultrasound, there are several research efforts on wearable and implantable ultrasound based devices for clinical applications \cite{la2022flexible, hu2023stretchable, liu2024wearable}. However, continuous evaluation of clinical metrics necessitates integrated computer vision algorithms to detect relevant features from ultrasound images. In this study, we explore the efficacy of deep learning models for automatic hematoma tracking and anatomical segmentation on our B-mode ultrasound dataset of 10,223 images of porcine spinal cords (N=25). To our knowledge, this is the first open source dataset showcasing the anatomy (dura, CSF, pia, spinal cord, vertebral structures) of both healthy and injured spinal cords. We benchmark the performance of several state-of-the-art object detection models to localize the site of injury (i.e., hematoma) within the ultrasound images, along with semantic segmentation algorithms for segmenting the spinal cord into its corresponding anatomy. In our analysis, we propose a new metric to evaluate how effectively each model can be deployed on wearable and implantable ultrasound devices to enable continuous monitoring of SCI. Finally, we explore the zero-shot generalization capabilities (i.e., model's ability to adapt to previously unseen data) of these semantic segmentation models by evaluating the trained porcine models on human spinal cord images and discuss methods to improve these models for human clinical translation. Our work serves as a pioneering effort to automate continuous diagnostics in SCI, providing avenues for more personalized and proactive treatment to improve clinical outcomes. With this released dataset, we hope to promote both methodology development for improved computer vision on medical datasets, along with clinical investigations and applications in the domain of SCI. 

\section*{Previous Works}
Achieving high accuracy and generalizability in supervised computer vision in spinal cord images typically requires training data that are sufficiently dense and representative of the domain -- a costly and time-consuming challenge in the medical field \cite{tajbakhsh2020embracing}. In contrast to the vast databases of natural images, medical datasets are often scarce and noisy, presenting a substantial challenge for supervised deep learning methods \cite{deng2009imagenet, lin2014microsoft, everingham2010pascal}. In this section, we highlight some recent efforts for developing medical datasets of spinal cord images. 
\par
Prados et al. provided axial MRI images of the cervical spine from 80 healthy subjects across 4 acquisition centers, with annotated white and gray matter segmentation masks \cite{prados2017spinal}.  Another published MRI dataset includes spinal cord images from 260 participants to study quantitative metrics (e.g., cross-sectional area of the cord and white/gray matter) of the regions of interest \cite{cohen2021open}. CTSpine1k provides 1,005 CT volumes with labeled vertebrae for bone segmentation across different spinal cord conditions \cite{deng2021ctspine1k}. However, to our knowledge there are no existing open-source datasets of spinal cord ultrasound images. Unlike CT and MRI, ultrasound has the capability to provide real-time continuous imaging for a unique perspective on spinal cord pulsatility and soft-tissue distinction, allowing clinical and surgical decision-making. 
\par 
There have been several studies which aim to use machine learning (ML) methods to automatically detect injury and segment the spinal cord. Research groups have proposed segment-based classification models \cite{ahammad2019fast, ahammad2020efficient} and hybrid thresholding and convolutional neural network (CNN)-based approaches to identify the severity of SCI and predict disease patterns using the segmented features in an SCI MR image database \cite{ahammad2020hybrid}. Other groups have used deep-learning based techniques for lesion (e.g., cervical disc herniation, traumatic SCI) and compression detection within MR images of the cervical spine \cite{ma2020faster, merali2021deep}.
\par 
Spinal cord segmentation efforts include Paugum et al.'s neural network framework \cite{paugam2019open} and a spinal cord segmentation challenge to segment white and gray matter in spinal cord structures from axial MRI data \cite{prados2017spinal}. Another CNN-based study proposed a spinal cord segmentation model on 2D axial-view MRI slices from 20 patients (359 images) with cervical spondylotic myelopathy \cite{zhang2021automatic}. 
\par
While there are many computer vision techniques for spinal cord segmentation and disease identification, the published methods at present are limited to MRIs and CTs, which do not provide continuous imaging or detailed soft-tissue delineation \cite{garg2021spinal}. As the deep learning field increasingly focuses on video-based understanding, leveraging consecutive frames to enhance network constraints, ultrasound stands out as the optimal modality for AI methods due to its high frame rate for data acquisition. The only other effort for segmentation in ultrasound spinal cord is done by Benjdira et al., where they evaluate models on images (N=10) post-laminectomy. In their analysis, they tested 3 ML networks on axial slices of the spinal cord to benchmark their performance on outputting a mask of the spinal cord \cite{benjdira2020spinal}. In our study, we aim to identify all relevant spinal cord anatomy (dura, CSF, pia, spinal cord) along with injury site within high resolution sagittal slices of the cord providing a longitudinal perspective of the spine for detecting conditions that extend over multiple vertebral levels. This dataset is publicly released to further the efforts for medical computer vision and automated injury monitoring. 

\section*{Methods}
\subsection*{Dataset formation}
\subsubsection*{Data acquisition} 
\begin{figure}[!b]
\centerline{\includegraphics[scale=0.5]{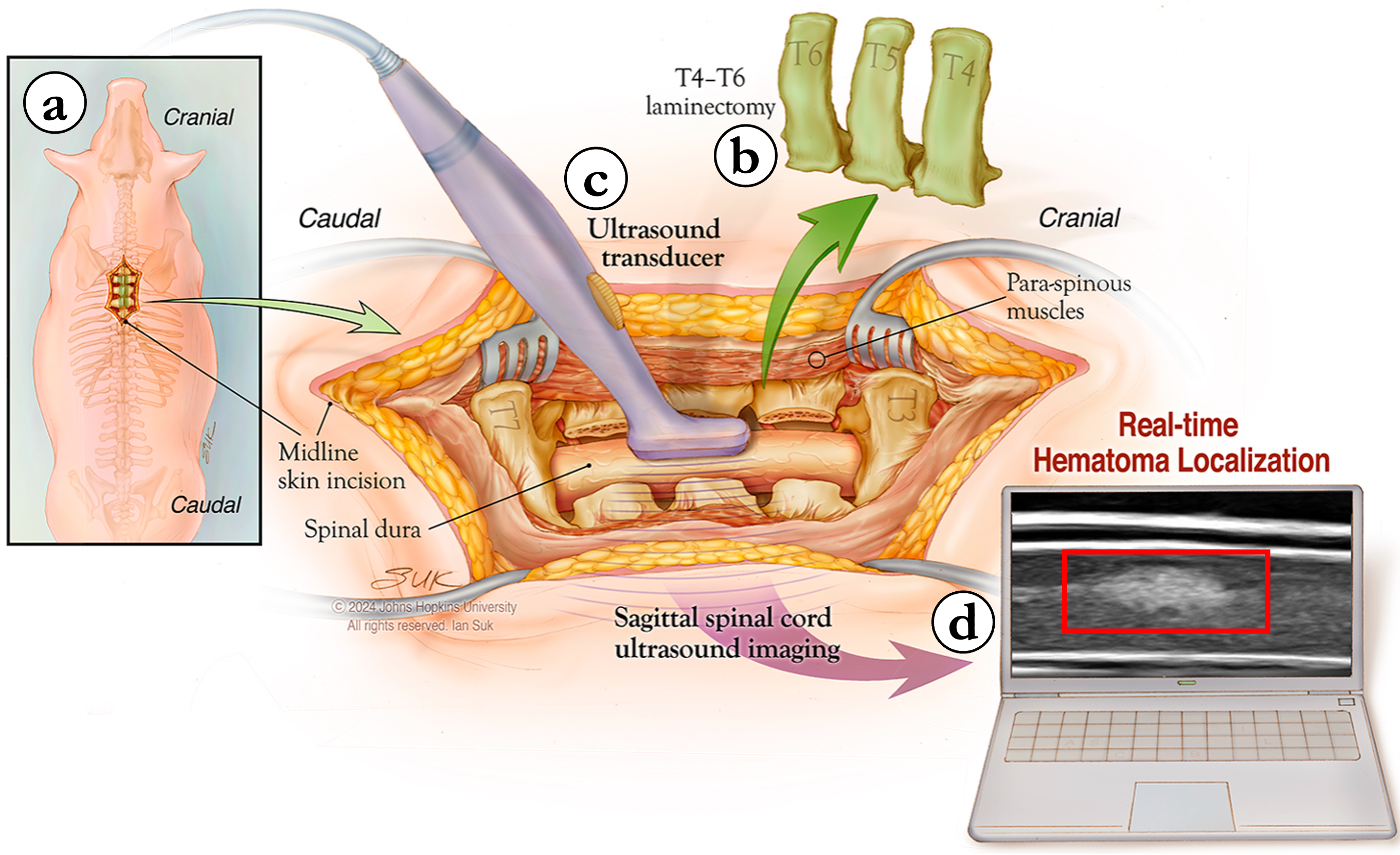}}\caption{Data collection of porcine spinal cord. (a) An aerial view of the female Yorkshire pig with a laminectomy to expose the spinal cord. (b) The spinous processes and lamina of the 4th to 6th thoracic vertebrae (T4-T6) are removed to provide an acoustic window and an injury is induced with a weight drop. (c) An i22LH8 transducer connected to Canon Aplio i800 ultrasound system is placed above the spinal cord to capture Brightness-mode (B-mode) images of the region of interest. (d) The resulting Digital Imaging and Communications in Medicine (DICOM) image is displayed on a personal computer, showcasing the dura, cerebrospinal fluid (CSF), pia, spinal cord, and the injury location (hematoma). The collected images are included in the final dataset for real-time injury localization and semantic segmentation.} 
\label{fig1}
\end{figure}

Over the course of two years (2021 - 2022), ultrasound data were collected from 25 female Yorkshire pigs weighing approximately 50 pounds. All animal experiments were conducted in compliance with the National Institutes of Health guide for the care and use of laboratory animals (NIH Publications No. 8023, revised 1978). All animal methods were approved by the Johns Hopkins University Animal Care and Use Committee (SW20M221) and were performed following the ARRIVE guidelines (Animal Research: Reporting of In Vivo Experiments). To collect the \textit{in vivo} ultrasound images in the dataset, a laminectomy was performed (Figure \ref{fig1}A) from the 4th to 6th thoracic vertebrae (T4 - T6) to provide an acoustic window (Figure \ref{fig1}B). Sagittal images were collected at different angles and locations of the spinal cord using a Canon Aplio i800 ultrasound system (Canon Medical Systems, Tustin, CA) connected to either an i22LH8 transducer (operating frequency: 20 MHz) or an i18LX5 transducer (operating frequency: 12 MHz) placed above the spinal cord (Figure \ref{fig1}C). A controlled injury was subsequently induced with a weight drop (either 20, 40, or 60 grams) from a height of 17 cm, and then images were recollected. Image acquisition parameters, such as spinal cord depth (distance of the spinal cord from the transducer), imaging angle, gain, brightness, and ultrasound signal coupling medium (saline solution), were varied to obtain an expressive and diverse dataset. These B-mode ultrasound images were collected by several different people during the data acquisition period, and were stored as Digital Imaging and Communications in Medicine (DICOM) files (the international standard for viewing and storing medical imaging information) for further preprocessing to make the final dataset (Figure \ref{fig1}D). 
\par
Additionally, B-mode images of the human spinal cord were captured from 8 patients after a laminectomy was performed from the 11th thoracic level to the 1st lumbar level (T11 - L1). Each patient had varying spinal cord curvatures, with 6 undergoing a posterior vertebral column subtraction osteotomy (PVCSO) \cite{theodore2021posterior}. Similar to the animal data collection, an i18LX5 transducer connected to a Canon Aplio i800 ultrasound system was placed above the spinal cord and images were collected throughout different stages of the shortening procedure \cite{kerensky2024tethered}. These images were used to evaluate the semantic segmentation models that were trained on only the porcine spinal cord images. All human data collection methods were approved by the Johns Hopkins Institutional Review Board (IRB00273900). All human data collection methods were performed in accordance with the Declaration of Helsinki and were acquired with informed consent.
\begin{figure*}[!b]
\centerline{\includegraphics[scale = 0.4]{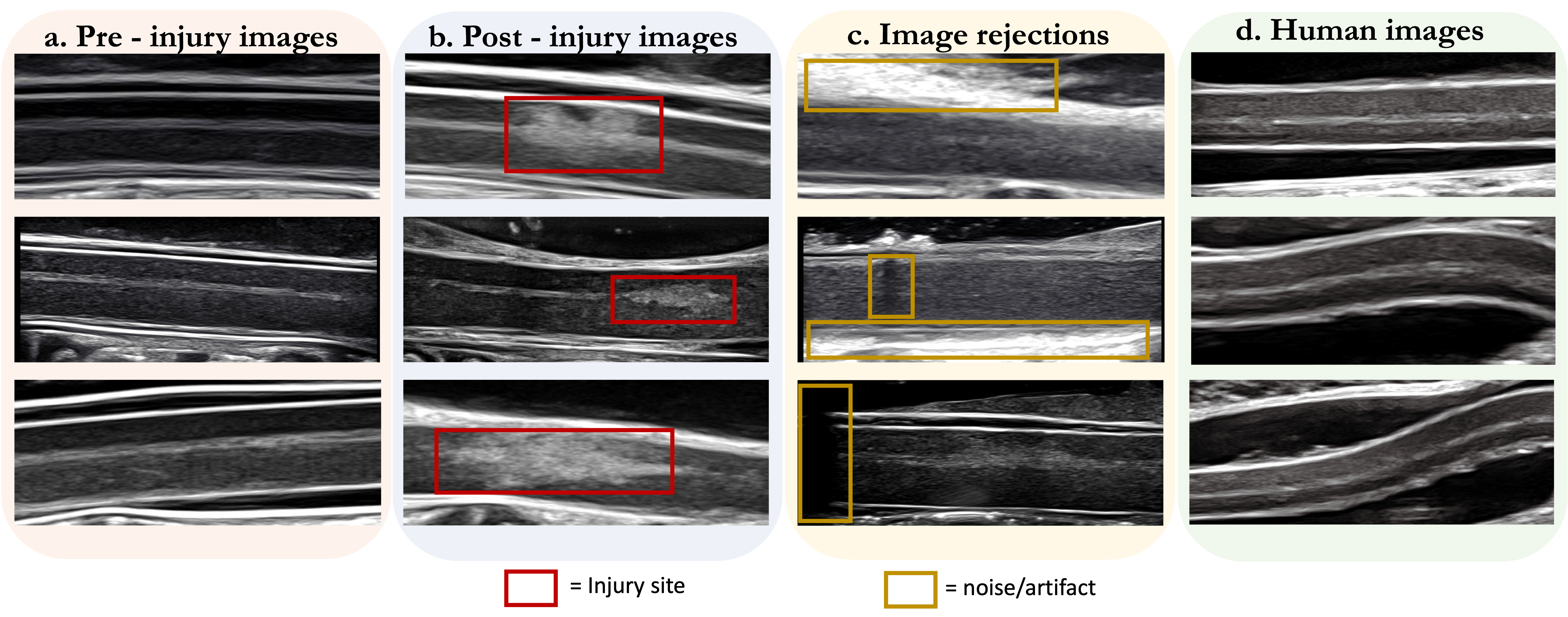}}
\caption{Curation of the porcine and human ultrasound spinal cord dataset. (a) Sample pre-injury images from the porcine spinal cord dataset. These images included the primary anatomy of interest and did not suffer from severe noise or artifacts that would render the image difficult to interpret. (b) Sample post-injury images of the dataset that fulfilled the same inclusion criteria as the pre-injury images. The red bounding boxes indicate the hematoma. (c) Sample images that were excluded from the final dataset due to shadowing, noise, or artifacts that affect image quality and occlude the anatomy of interest. (d) Sample human spinal cord images that were used to test the generalizability of the segmentation models.} 
\label{fig2}
\end{figure*}

\subsubsection*{Data description} 
There were a total of 3 datasets acquired and created by our group as part of this study for: (1) injury localization in porcine images, (2) semantic segmentation in porcine images, (3) semantic segmentation in human images. After rejecting low quality images in which the spinal cord was occluded or corrupted with noise, the injury localization dataset was developed by randomly selecting 2,245 healthy and injured porcine spinal cord images from the collected data (N=23). The semantic segmentation dataset consists of 10,223 images divided into 4,467 healthy spinal cord images (N=20) and 5,756 images of injured spinal cord images (N=25). To reduce file size, the images (1280 × 960 pixels) were converted to a Portable Network Graphics (PNG) format, scaled to the depth of scan field of the ultrasound image, and cropped to include only the anatomy of interest (690 × 275 pixels) (Figure \ref{fig2}A-C). After preprocessing, each image shows 25 mm × 8 mm of the sagittal slice of the spinal cord. Finally, a third dataset consisting of 86 human spinal cord ultrasound images (N=8) was created and preprocessed to develop a unique test set for the semantic segmentation models. While these images did not include a hematoma within the cord, they showed the other anatomy of interest (i.e., dura, CSF, pia, spinal cord), as shown in Figure \ref{fig2}D. 

\subsubsection*{Data annotation}
\begin{figure}[!t]
\centerline{\includegraphics[scale=0.4]{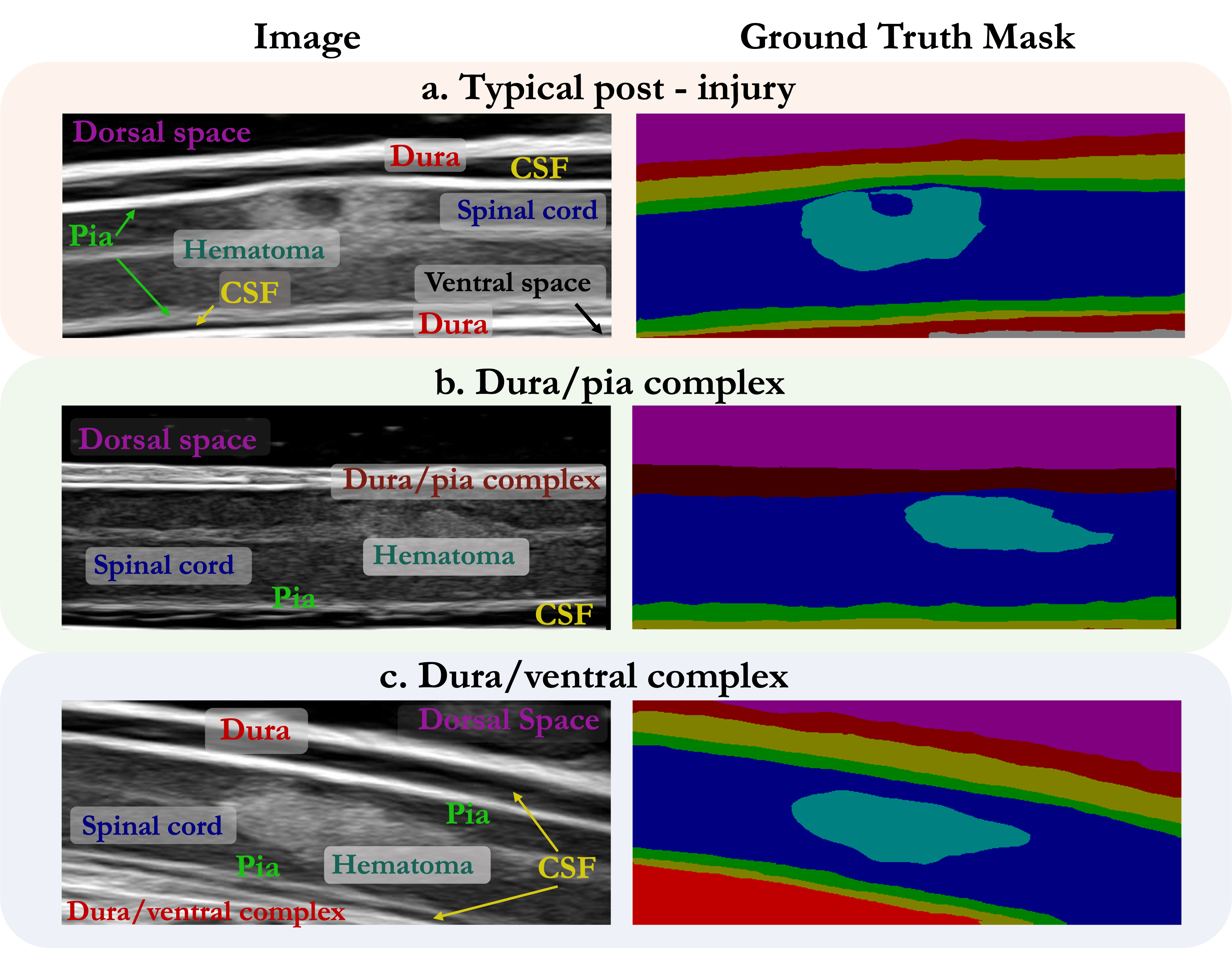}}
\caption{Example images and their corresponding ground truth masks. (a) A typical spinal cord image displaying clear delineation between the anatomical boundaries. The bottom of the image corresponds to the ventral spinal cord. (b) A spinal cord image in which the induced injury caused swelling in the tissue, effectively blurring the delineation between the dura, cerebrospinal fluid (CSF), and the pia. To avoid mistakes in interpretation and inconsistencies in labelling, the region is annotated as the dura/pia complex. (c) In some images, noise and other intraoperative artifacts resulted in ambiguous delineation between the ventral dura and the ventral region. For these types of images, we label that anatomy as the dura/ventral complex.} 
\label{fig3}
\end{figure}

The injury localization and anatomical segmentation tasks had different annotation processes. For injury localization, images with a hematoma present were labeled with a bounding box surrounding the injury. The ground truth image masks used for semantic segmentation were created in Computer Vision Annotation Tool (CVAT) (CVAT.ai Corporation, Palo Alto, CA). The anatomical structures of interest were the dura, CSF, pia, spinal cord, and hematoma (Figure \ref{fig3}A). Additionally, the space above the dorsal dura, where the vertebral bone was removed, was labeled as the dorsal space, and the space below the ventral dura consisting of bony structures and cartilage was labeled as the ventral space. For some injuries, depending on the angle in which the ultrasound images were captured and the impact of the weight drop, the anatomical boundaries between the dura, CSF, and pia were unclear and could not be labeled with sufficient confidence. In order to maintain the integrity of the labels and minimize speculations during the labeling process, those regions were simplified to the dura/pia complex whenever necessary (Figure \ref{fig3}B). Similarly, the dura/ventral complex label was used when there was ambiguity in the anatomical boundary between the ventral dura and the vertebral structures beneath it (Figure \ref{fig3}C). The third dataset of human images was labeled using the same annotation process described above for the porcine spinal cord.
\par
Images were labeled by a team of medical and graduate students trained by a board-certified radiologist. Each image in which there was a lack of clear delineation in anatomical boundaries was verified by the radiologist to ensure high quality and accurate labels. Once all the images were labeled, the masks were then validated by a neurosurgery spine fellow at the Johns Hopkins Hospital to provide a second round of verification and further ensure the robustness of the labels. Table \ref{tab:dataset} provides an overview of the image and pixel instances for each of these labels. These images and their corresponding annotations are available electronically on our \href{https://github.com/HEPIUSLAB/ultrasound_spinal_cord_dataset.git}{\underline{GitHub Respository}}.

\subsection*{Injury Localization}
Using the injury localization dataset of 2,245 images, we evaluated the performance of seven state-of-the-art object detection models:  Faster RCNN \cite{girshick2015fast}, SSD300, SSD512 \cite{liu2016ssd}, RetinaNet \cite{lin2017focal}, Detection Transformer (DETR) \cite{carion2020end}, YOLOv7 \cite{wang2023yolov7}, and YOLOv8 \cite{Jocher_Ultralytics_YOLO_2023}. Faster RCNN is a two-stage detector where the first stage proposes candidate object bounding boxes with a Region Proposal Network, followed by a second stage that classifies the object and refines its bounding box \cite{girshick2015fast}. Conversely, the SSD model divides the image into a grid and predicts bounding boxes and class probabilities for each grid cell, with SSD300 optimizing for speed with 300 × 300 pixel images, and SSD512 catering to higher-resolution 512 × 512 pixel images \cite{liu2016ssd}. RetinaNet uses a Feature Pyramid Network for building high-level semantic feature maps at multiple scales to efficiently detect objects of various sizes and introduces the Focal Loss function to address the challenge of foreground-background class imbalance \cite{lin2017focal}. In contrast, DETR uniquely combines a CNN backbone with a transformer encoder-decoder architecture, employing learned positional embeddings and a shared feed-forward network for precise object class and bounding box predictions \cite{carion2020end}. YOLOv7, utilizes the Extended Efficient Layer Aggregation Network (E-ELAN) to optimize convergence and network efficiency \cite{wang2023yolov7}. Lastly, YOLOv8 utilizes an anchor-free architecture, enhancing detection accuracy across various object sizes and shapes, complemented by a versatile multi-scale prediction method for improved adaptability \cite{Jocher_Ultralytics_YOLO_2023}. More specific details on the models’ architectures can be found in their respective references. These models have exhibited high performance when applied to natural image datasets, like COCO \cite{lin2014microsoft}, PASCAL VOC \cite{everingham2010pascal}, and ImageNet \cite{deng2009imagenet}. The learnable parameters of the model were initialized with pretrained weights and biases using the ImageNet dataset to reduce the training cost.

The injury localization dataset was divided into 3 splits with roughly 80\% of both pre- and post-injury images for training, 10\% for validation, and 10\% for testing. We ensured that all images taken from each subject were placed in the same group to prevent overfitting during training, and we balanced the distribution of healthy and injured spinal cord images in each subset. To improve the robustness of the models, we performed a number of data augmentations in the training process. This includes image transformations (e.g., rotating the samples with a degree range of $\pm$ 25, horizontal flipping with a probability of 0.5, and randomly zooming in or out of the image by 30\%) and image degradation (e.g., blurring the image by a factor of 3 with a probability of 20\%). We also adjusted image appearance by randomly manipulating the statistical characteristics of the image intensities such as brightness, saturation, hue, and contrast by a factor of $\pm$ 0.2. To optimize computational resources and time, object detection models uniformly resize images to dimensions of 256 x 256 pixels. These models were trained on a Windows 11 Machine (8 GB RAM) with 24 GB NVIDIA GeForce RTX 3090 graphics unit and 14th Gen Intel Core i5-14600 processor (14 cores, 20 threads, 2.7GHZ to 5.2GHz turbo frequency). The hyperparameters of the model (i.e., batch size, learning rate, training epochs) were tuned to ensure optimal performance using the Neural Network Intelligence software (Supplementary Table 1-7) \cite{nni2021}. The main characteristics of each model are described in Table \ref{tab:object-detection-description}. 
\par
\begin{table*}[htb]
\centering
\caption{Pixel- and image-wise instances of each anatomical structure.}
\label{tab:dataset}
\begin{tabular}{|c|c|c|}
\hline
\textbf{Anatomy}     & \textbf{Pixel Instances} & \textbf{Instances across Images} \\ \hline
Dorsal Space         & 397,350,611              & 10,223                           \\ \hline
Dura                 & 142,857,955              & 9,814                            \\ \hline
Pia                  & 142,721,894              & 9,839                            \\ \hline
CSF                  & 118,161,757              & 9,613                            \\ \hline
Dura/Pia complex     & 57,124,376               & 2,732                            \\ \hline
Spinal cord          & 812,894,511              & 10,223                           \\ \hline
Hematoma             & 69,727,644               & 5,756                            \\ \hline
Dura/Ventral Complex & 58,152,053               & 2,671                            \\ \hline
Ventral Space        & 89,571,059               & 6,099                            \\ \hline
Background           & 18,992,200               & 6,385                            \\ \hline
\end{tabular}
\end{table*}
\begin{table*}[htb]
\centering
\caption{Characteristics of object detection models to detect the site of injury on porcine spinal cord ultrasound images.}

\scalebox{0.9}{
\label{tab:object-detection-description}
\begin{tabular}{|c|c|c|c|c|c|c|c|}
\hline
\textbf{Model} & \multicolumn{1}{l|}{\textbf{Encoder}} & \textbf{\begin{tabular}[c]{@{}c@{}}\# of \\ Parameters\end{tabular}} & \textbf{\begin{tabular}[c]{@{}c@{}}Learning\\ Rate\end{tabular}} & \textbf{\begin{tabular}[c]{@{}c@{}}Batch\\ Size\end{tabular}} & \textbf{\begin{tabular}[c]{@{}c@{}}Training \\ Epochs\end{tabular}} & \textbf{Optimizer} & \textbf{Loss Function} \\
\hline
\begin{tabular}[c]{@{}c@{}}Faster \\ RCNN\end{tabular} & ResNet50 & 41,299,161 & 0.004935  & 4 & 60 & SGD & \begin{tabular}[c]{@{}c@{}}Cross Entropy + Smooth L1\end{tabular} \\
\hline
SSD300 & ResNet50 & 24,641,780 & 0.002571 & 32 & 20 & SGD & \begin{tabular}[c]{@{}c@{}}Cross Entropy + Smooth L1\end{tabular} \\
\hline
SSD512 & ResNet50 & 24,641,780 & 0.000251 & 8 & 20 & SGD & \begin{tabular}[c]{@{}c@{}}Cross Entropy + Smooth L1\end{tabular} \\
\hline
RetinaNet & ResNet50 & 36,352,630 & 0.000099 & 8 & 60 & SGD & Focal Loss \\
\hline
DETR & ResNet50 & 41,524,954 & 0.000011 & 8 & 200 & AdamW & \begin{tabular}[c]{@{}c@{}}Cross Entropy + Smooth L1\end{tabular} \\
\hline
YOLOv7 & E-ELAN & 37,196,556 & 0.000123  & 32 & 75 & Adam & \begin{tabular}[c]{@{}c@{}}Binary Cross Entropy + Mean Square Error\end{tabular} \\
\hline
YOLOv8 & CSPNet & 25,856,899 & 0.000422 & 48 & 80 & \begin{tabular}[c]{@{}c@{}}AdamW\\ + SGD\end{tabular} & \begin{tabular}[c]{@{}c@{}}Binary Cross Entropy + Distribution Focal \\ Loss + Complete Intersection over Union\end{tabular} \\
\hline
\end{tabular}
}
\end{table*}

\subsection*{Spinal Cord Segmentation}
We evaluated six state-of-the-art semantic segmentation models on our dataset consisting of 10,223 porcine images annotated with masks delineating the anatomical structures.  The specific models used were SegFormer \cite{xie2021segformer}, U-Net \cite{ronneberger2015u}, DeepLabv3 \cite{deeplabv32018}, TransUNet \cite{chen2021transunet}, Swin-UNet \cite{cao2022swin}, and Segment Anything for medical images (SAMed) \cite{samed}. SegFormer combines a transformer-based encoder to process global information and generate multiscale features, with a straightforward multilayer perceptron (MLP) decoder that integrates these features at various scales  \cite{xie2021segformer}. U-Net, renowned for its effectiveness in biomedical imaging, employs a contracting-expanding architecture of the convolutional layers (resulting in a u-shape) that balances spatial and feature information to ensure accurate localization \cite{ronneberger2015u}. DeepLabv3 enhances multi-scale context capture through an Atrous Spatial Pyramid Pooling (ASPP) module, which uses dilated convolutions at various scales \cite{deeplabv32018}. TransUNet leverages the global self-attention mechanisms of Transformers to augment the traditional U-Net’s convolutional approach, offering improved global information processing and localization \cite{chen2021transunet}. Swin-UNet adopts a Swin Transformer in both its encoder and decoder, utilizing shifted windows to enhance context feature extraction and spatial resolution restoration \cite{cao2022swin}. Lastly, SAMed builds on the Segment Anything framework, specifically adapting its encoder and introducing a low-rank-based fine-tuning approach for medical image segmentation \cite{samed}.
\par
The segmentation dataset was again randomly subdivided into a roughly 80-10-10 split for training, validation, and testing, respectively. All images taken from the same porcine spine were included in a single subset (train, validation, or test) to avoid overlap and prevent artificially inflated results. Additionally, each subset contained an approximately equal number of healthy and injured spinal cords. The training dataset was augmented to improve the models' generalizability with image transformations (e.g., rotation [$\pm$ 50-degree limit], horizontal flipping [p = 0.5], image compression [p = 0.2]), and image degradation (random brightness contrast [p = 0.2], random fog [p = 0.2]). For computational and time efficiency, all segmentation models resize the images to 256 × 256 pixels. The hyperparameters (i.e., batch size, learning rate, training epochs) were tuned in the same manner as the object detectors, using the same Windows machine for training (Supplementary Table 8-13) \cite{nni2021}. Table \ref{tab:segmentation-description} describes the main characteristics of each semantic segmentation model. 
\par
\begin{table*}[htb]
\centering
\caption{Characteristics of semantic segmentation models for segmenting porcine spinal cord ultrasound images.}
\scalebox{0.95}{
\label{tab:segmentation-description}
\begin{tabular}{|c|c|c|c|c|c|c|c|}
\hline
\textbf{Model} & \textbf{Encoder} & \textbf{\begin{tabular}[c]{@{}c@{}}\# of \\ Parameters\end{tabular}} & \textbf{\begin{tabular}[c]{@{}c@{}}Learning\\ Rate\end{tabular}} & \textbf{\begin{tabular}[c]{@{}c@{}}Batch\\ Size\end{tabular}} & \textbf{\begin{tabular}[c]{@{}c@{}}Training\\ Epochs\end{tabular}} & \textbf{Optimizer} & \textbf{Loss Function} \\
\hline
SegFormer & MiT-B5 & 84,601,034 & 0.000972 & 4 & 75 & AdamW & Cross Entropy \\
\hline
U-Net & ResNet50 & 31,044,106 & 0.003011 & 8 & 50 & SGD & Cross Entropy \\
\hline
DeepLabv3 & ResNet50 & 41,998,420 & 0.000334 & 4 & 100 & AdamW & Cross Entropy \\
\hline
TransUNet & \begin{tabular}[c]{@{}c@{}}ResNet50 + ViT\_B16\end{tabular} & 105,323,306 & 0.004468 & 24 & 200 & SGD & \begin{tabular}[c]{@{}c@{}}Cross Entropy + Dice\end{tabular} \\
\hline
Swin-UNet & Swin-T & 27,153,156 & 0.061411 & 16 & 100 & SGD & \begin{tabular}[c]{@{}c@{}}Cross Entropy + Dice\end{tabular} \\
\hline
SAMed & SAM ViT-B & 91,866,903 & 0.002840 & 24 & 200 & AdamW & \begin{tabular}[c]{@{}c@{}}Cross Entropy + Dice\end{tabular} \\
\hline
\end{tabular}
}
\end{table*}
Because the porcine and human spinal cord have similar anatomical structures and immune response after injury, we evaluated the semantic segmentation models on human spinal cord ultrasound images to understand whether models trained exclusively on porcine images can generalize with sufficient accuracy to human images \cite{toossi2021comparative}. The annotated dataset containing the 86 human spinal cord images (N=8 patients) were used as the test set for the trained models to highlight the translatability of our trained models to clinical application. 

\section*{Results}
\subsection*{Injury Localization}
After hyperparameter tuning, the performance of each model during inference is shown in Table \ref{tab:object-detection-results}, including the mean Average Precision (mAP) value, the Average Recall (AR) value, and speed using frames per second (FPS). mAP is a popular metric ranging from 0 to 1 that measures the accuracy of object detectors as it provides a comprehensive assessment of their performance, considering both precision (how accurate the detected objected are) and recall (how many relevant objects are detected) at multiple levels of confidence (i.e., 50\%, and 50-95\% at intervals of 5\%). The levels of confidence, known as the Intersection over Union (IoU) thresholds, consider a detection accurate if the IoU of the ground truth bounding box and the predicted bounding box is greater than the set threshold (e.g., 0.5 for mAP50). The AR value is the recall metric averaged over a range of IoU threshold (0.5 - 1.0). From these results, it is evident that Faster RCNN and YOLOv8 show the strongest performance, achieving a mAP50 score of 0.985 and 0.979, respectively. These models also achieve the highest mAP50-95 score, which is a much more stringent metric for assessing model performance compared to mAP50, at 0.524 for Faster RCNN and 0.606 for YOLOv8. YOLOv8 also attains the highest AR score at 0.644.
\par 
\begin{table*}
\centering
\caption{Performance of object detection models on unseen porcine spinal cord ultrasound images to detect the site of injury.}
\scalebox{0.8}{
\label{tab:object-detection-results}
\begin{tabular}{|c|c|c|c|ccc|ccc|}
\hline
\multirow{2}{*}{\textbf{Model}} &
  \multirow{2}{*}{\textbf{mAP50}} &
  \multirow{2}{*}{\textbf{mAP50-95}} &
  \multirow{2}{*}{\textbf{AR}} &
  \multicolumn{3}{c|}{\textbf{CPU}} &
  \multicolumn{3}{c|}{\textbf{GPU}} \\ \cline{5-10} 
 &
   &
   &
   &
  \multicolumn{1}{c|}{\textbf{FPS}} &
  \multicolumn{1}{c|}{\textbf{Load (\%)}} &
  \textbf{\begin{tabular}[c]{@{}c@{}}Implantability\\ Score\end{tabular}} &
  \multicolumn{1}{c|}{\textbf{FPS}} &
  \multicolumn{1}{l|}{\textbf{Load (\%)}} &
  \textbf{\begin{tabular}[c]{@{}c@{}}Implantability \\ Score\end{tabular}} \\ \hline
\begin{tabular}[c]{@{}c@{}}Faster RCNN\end{tabular} &
  \textbf{0.985} &
  0.524 &
  0.594 &
  \multicolumn{1}{c|}{1.53} &
  \multicolumn{1}{c|}{33} &
  0.676 &
  \multicolumn{1}{c|}{14.98} &
  \multicolumn{1}{c|}{62} &
  0.613 \\ \hline
SSD300 &
  0.669 &
  0.207 &
  0.249 &
  \multicolumn{1}{c|}{23.12} &
  \multicolumn{1}{c|}{36} &
  0.735 &
  \multicolumn{1}{c|}{\textbf{147.62}} &
  \multicolumn{1}{c|}{27} &
  0.766 \\ \hline
SSD512 &
  0.874 &
  0.274 &
  0.324 &
  \multicolumn{1}{c|}{\textbf{23.46}} &
  \multicolumn{1}{c|}{26} &
  0.866 &
  \multicolumn{1}{c|}{125.03} &
  \multicolumn{1}{c|}{18} &
  0.853 \\ \hline
RetinaNet &
  0.912 &
  0.264 &
  0.426 &
  \multicolumn{1}{c|}{1.89} &
  \multicolumn{1}{c|}{32} &
  0.646 &
  \multicolumn{1}{l|}{17.66} &
  \multicolumn{1}{c|}{48} &
  0.616 \\ \hline
DETR &
  0.787 &
  0.251 &
  0.453 &
  \multicolumn{1}{c|}{19.78} &
  \multicolumn{1}{c|}{15} &
  0.812 &
  \multicolumn{1}{c|}{114.35} &
  \multicolumn{1}{c|}{26} &
  0.772 \\ \hline
YOLOv7 &
  0.923 &
  0.439 &
  0.499 &
  \multicolumn{1}{c|}{19.54} &
  \multicolumn{1}{c|}{20} &
  0.865 &
  \multicolumn{1}{c|}{80.13} &
  \multicolumn{1}{c|}{28} &
  0.777 \\ \hline
YOLOv8 &
  0.979 &
  \textbf{0.606} &
  \textbf{0.644} &
  \multicolumn{1}{c|}{17.85} &
  \multicolumn{1}{c|}{22} &
  \textbf{0.870} &
  \multicolumn{1}{l|}{115.31} &
  \multicolumn{1}{c|}{27} &
  \textbf{0.867} \\ \hline
\end{tabular}
}
\end{table*}
In an effort to better assess the potential of incorporating these models into implantable or wearable devices, we propose a new metric, implantability score, that takes into account the model accuracy, speed, and computational load during inference. We evaluate this implantability score on both the CPU and the GPU with a weighted average of the mAP50 score, normalized average FPS, and average processing load, as shown in Equation (\ref{implant}).    
\begin{equation}
\textit{Implantability Score} = \frac{\textit{mAP50}}{2} +  \frac{\textit{FPS}_{\textit{norm}}}{4} + \frac{1-\textit{Load}}{4}
\label{implant}
\end{equation}
The mAP50 score is an accuracy metric between 0 and 1. This is weighted twice as much as the other metrics to determine the implantability score as this has a significant impact on the diagnostic capability of the implant. The FPS is normalized to a metric between 0 and 1 by dividing all results with an FPS\textsubscript{max} value. The FPS\textsubscript{max} is set to the highest recorded FPS by any model on our CPU (24 FPS). To adapt the FPS normalization process for GPU applications, which generally have much higher speed, the FPS\textsubscript{max} is set to the highest speed across all the evaluated models (148 FPS). Finally, because lower computational load is more desirable in this context, we use the the inverted average CPU or GPU load (1-Load), which is the percentage increase of computational power during inference, for calculating the score. Our results indicate that YOLOv8 has optimal characteristics for injury localization for continuous monitoring with ultrasound-based implants for both CPU and GPU based applications, with a CPU implantability score of 0.870 and GPU implantability score of 0.867.  

\subsection*{Spinal Cord Segmentation}
After hyperparameter tuning, the performance of each segmentation model during inference is shown in Table \ref{tab:segmentation-results}, including the Mean Intersection over Union (MIoU) value, speed, and computational load. MIoU is a measure of the overlap between the predicted segmentation and ground truth segmentation. The IoU for a single class (i.e., each anatomical structure) is calculated as the intersection of the predicted and true positive pixels divided by the union of predicted and true positive pixels and the MIoU is the average IoU across all classes. This provides a more comprehensive understanding of how well the model is performing across different classes. The Mean Dice coefficient, which is another metric between 0 and 1 that indicates the level of similarity between the predicted and ground truth masks, is also evaluated. It is calculated as two times the area of the intersection of the predicted and ground truth masks divided by the sum of the areas of the predicted and ground truth mask.
\par
\begin{figure*}[!ht]
\centerline{\includegraphics[scale=0.4]{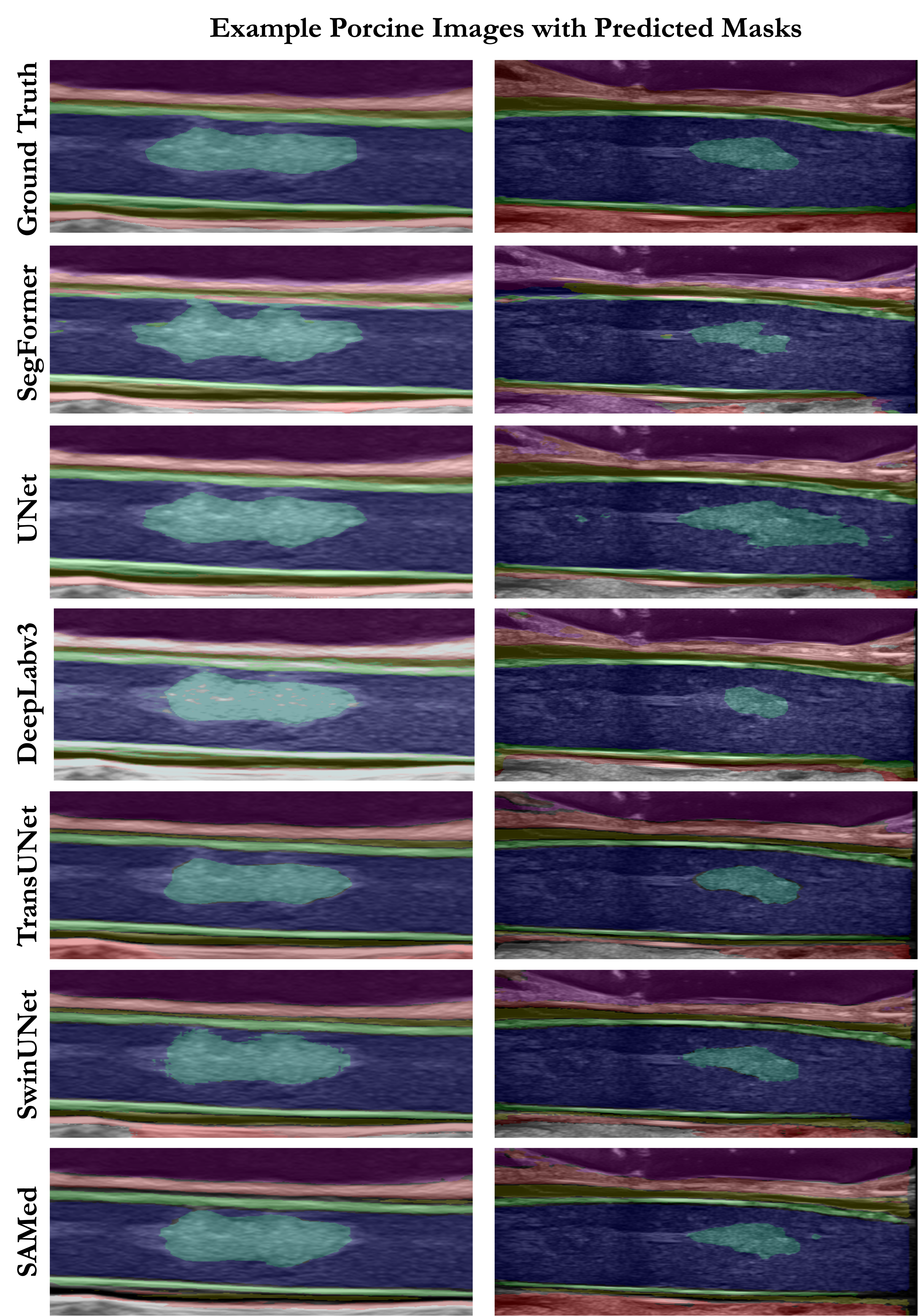}}
\caption{Visualization of the semantic segmentation models' performance overlaid on example porcine images.} 
\label{fig4}
\end{figure*}
The class-wise IoU scores and Dice scores are included in the supplementary material (Supplementary Table 14 - 15). The metrics are evaluated on both unseen porcine images and human images, which the model had not been exposed to during the training process. To assess the potential for clinical translation, the zero-shot generalizability of these models on human spinal cord images is measured, including the MIoU and Mean Dice scores for the entire anatomy, along with class-wise scores (Supplementary Table 16 - 18). The implantable score is again computed using Equation (\ref{implant}), replacing mAP50 with Mean Dice.
\par
Our results indicate that DeepLabv3 outperforms all other segmentation models in terms of accuracy on porcine anatomy, with a Mean Dice score of 0.587, and SAMed generalizes best to human anatomy, achieving a Mean Dice score of 0.445. TransUNet generalizes the best to human spinal cord, with a Dice coefficient of 0.853 for the spinal cord class. Taking accuracy, speed, and computational load into account to determine the model's implantable score (i.e., potential for deployment on an implantable or wearable device), SwinUNet outperforms all other models for CPU-based chips with an implantability score of 0.699 and DeepLabv3 achieves the highest score for GPU-based devices at 0.702. Figure \ref{fig4} depicts the predicted segmentation masks for each model on an unseen porcine spinal cord image, along with the original ground truth image and mask.
\begin{table*}[htb]
\centering
\caption{Performance of semantic segmentation models on unseen porcine and human spinal cord ultrasound images.}
\scalebox{0.8}{
\label{tab:segmentation-results}
\begin{tabular}{|c|cc|cc|cc|cc|ccc|ccc|}
\hline
\multirow{2}{*}{\textbf{Model}} &
  \multicolumn{2}{c|}{\textbf{\begin{tabular}[c]{@{}c@{}}Porcine\\ anatomy\end{tabular}}} &
  \multicolumn{2}{c|}{\textbf{\begin{tabular}[c]{@{}c@{}}Porcine\\ spinal cord\end{tabular}}} &
  \multicolumn{2}{c|}{\textbf{\begin{tabular}[c]{@{}c@{}}Human\\ anatomy\end{tabular}}} &
  \multicolumn{2}{c|}{\textbf{\begin{tabular}[c]{@{}c@{}}Human \\ spinal cord\end{tabular}}} &
  \multicolumn{3}{c|}{\textbf{CPU}} &
  \multicolumn{3}{c|}{\textbf{GPU}} \\ \cline{2-15} 
 &
  \multicolumn{1}{c|}{\textbf{MIoU}} &
  \textbf{Dice} &
  \multicolumn{1}{c|}{\textbf{IoU}} &
  \textbf{Dice} &
  \multicolumn{1}{c|}{\textbf{MIoU}} &
  \textbf{Dice} &
  \multicolumn{1}{c|}{\textbf{IoU}} &
  \textbf{Dice} &
  \multicolumn{1}{c|}{\textbf{FPS}} &
  \multicolumn{1}{c|}{\textbf{\begin{tabular}[c]{@{}c@{}}Load \\ (\%)\end{tabular}}} &
  \textbf{\begin{tabular}[c]{@{}c@{}}Implantability\\ Score\end{tabular}} &
  \multicolumn{1}{c|}{\textbf{FPS}} &
  \multicolumn{1}{c|}{\textbf{\begin{tabular}[c]{@{}c@{}}Load \\ (\%)\end{tabular}}} &
  \textbf{\begin{tabular}[c]{@{}c@{}}Implantability\\ Score\end{tabular}} \\ \hline
SegFormer &
  \multicolumn{1}{c|}{0.493} &
  0.570 &
  \multicolumn{1}{c|}{0.906} &
  0.950 &
  \multicolumn{1}{c|}{0.232} &
  0.308 &
  \multicolumn{1}{c|}{0.666} &
  0.773 &
  \multicolumn{1}{c|}{3.66} &
  \multicolumn{1}{c|}{23} &
  0.548 &
  \multicolumn{1}{c|}{23.50} &
  \multicolumn{1}{c|}{45} &
  0.513 \\ \hline
U-Net &
  \multicolumn{1}{c|}{0.476} &
  0.553 &
  \multicolumn{1}{c|}{0.867} &
  0.928 &
  \multicolumn{1}{c|}{0.253} &
  0.349 &
  \multicolumn{1}{c|}{0.609} &
  0.722 &
  \multicolumn{1}{c|}{6.06} &
  \multicolumn{1}{c|}{32} &
  0.563 &
  \multicolumn{1}{c|}{62.37} &
  \multicolumn{1}{c|}{41} &
  0.668 \\ \hline
DeepLabv3 &
  \multicolumn{1}{c|}{\textbf{0.515}} &
  \textbf{0.587} &
  \multicolumn{1}{c|}{0.910} &
  0.952 &
  \multicolumn{1}{c|}{0.200} &
  0.289 &
  \multicolumn{1}{c|}{0.506} &
  0.656 &
  \multicolumn{1}{c|}{4.76} &
  \multicolumn{1}{c|}{27} &
  0.568 &
  \multicolumn{1}{c|}{64.11} &
  \multicolumn{1}{c|}{35} &
  \textbf{0.702} \\ \hline
TransUNet &
  \multicolumn{1}{c|}{0.500} &
  0.573 &
  \multicolumn{1}{c|}{\textbf{0.921}} &
  \textbf{0.958} &
  \multicolumn{1}{c|}{0.298} &
  0.388 &
  \multicolumn{1}{c|}{\textbf{0.758}} &
  \textbf{0.853} &
  \multicolumn{1}{c|}{4.33} &
  \multicolumn{1}{c|}{35} &
  0.532 &
  \multicolumn{1}{c|}{40.85} &
  \multicolumn{1}{c|}{34} &
  0.609 \\ \hline
SwinUNet &
  \multicolumn{1}{c|}{0.490} &
  0.562 &
  \multicolumn{1}{c|}{0.913} &
  0.954 &
  \multicolumn{1}{c|}{0.309} &
  0.401 &
  \multicolumn{1}{c|}{0.692} &
  0.783 &
  \multicolumn{1}{c|}{12.75} &
  \multicolumn{1}{c|}{31} &
  \textbf{0.699} &
  \multicolumn{1}{c|}{63.36} &
  \multicolumn{1}{c|}{34} &
  0.690 \\ \hline
SAMed &
  \multicolumn{1}{c|}{0.497} &
  0.574 &
  \multicolumn{1}{c|}{0.908} &
  0.951 &
  \multicolumn{1}{c|}{\textbf{0.347}} &
  \textbf{0.445} &
  \multicolumn{1}{c|}{0.616} &
  0.740 &
  \multicolumn{1}{c|}{5.40} &
  \multicolumn{1}{c|}{37} &
  0.535 &
  \multicolumn{1}{c|}{29.43} &
  \multicolumn{1}{c|}{35} &
  0.563 \\ \hline
\end{tabular}
}
\end{table*}

\section*{Discussion}
Deep learning is a promising tool for medical image processing for computer-aided diagnostics in SCI. However, due to the lack of medical image data and their corresponding semantic labels, large-scale computer vision models have been underutilized in this context. With this study, we establish an avenue for automatic and continuous monitoring in SCI with our large-scale ultrasound dataset. After benchmarking several object detection and semantic segmentation models, our results show that YOLOv8 and DeepLabv3 are best suited for hematoma tracking and anatomical segmentation, respectively. While it can be challenging to understand the exact reason why some models outperform others, it can be expected that YOLOv8 exhibits the highest performance due to a few distinct characteristics. YOLOv8 incorporates advanced feature aggregation techniques that help in combining features from different scales. This multi-scale approach is particularly useful in medical imaging where the size and shape of the injury can vary significantly. Moreover, YOLOv8 uses mosaic data augmentation, which mixes 4 images together, to provide the model with improved contextual information. Additionally, it is anticipated that YOLOv8’s anchor-free detection scheme, in which the model directly predicts an object’s mid-point, improves generalization and inference speed for our custom dataset. For semantic segmentation, DeepLabv3 may exhibit the highest accuracy on our dataset because of the atrous convolution approach along with atrous spatial pyramid pooling for capturing multi-scale contextual information without losing resolution. This architecture enables DeepLabv3 to effectively handle boundary information and delineate object boundaries more accurately, which is crucial for our soft-tissue segmentation task.
\par
\begin{figure}
\centerline{\includegraphics[scale=0.37]{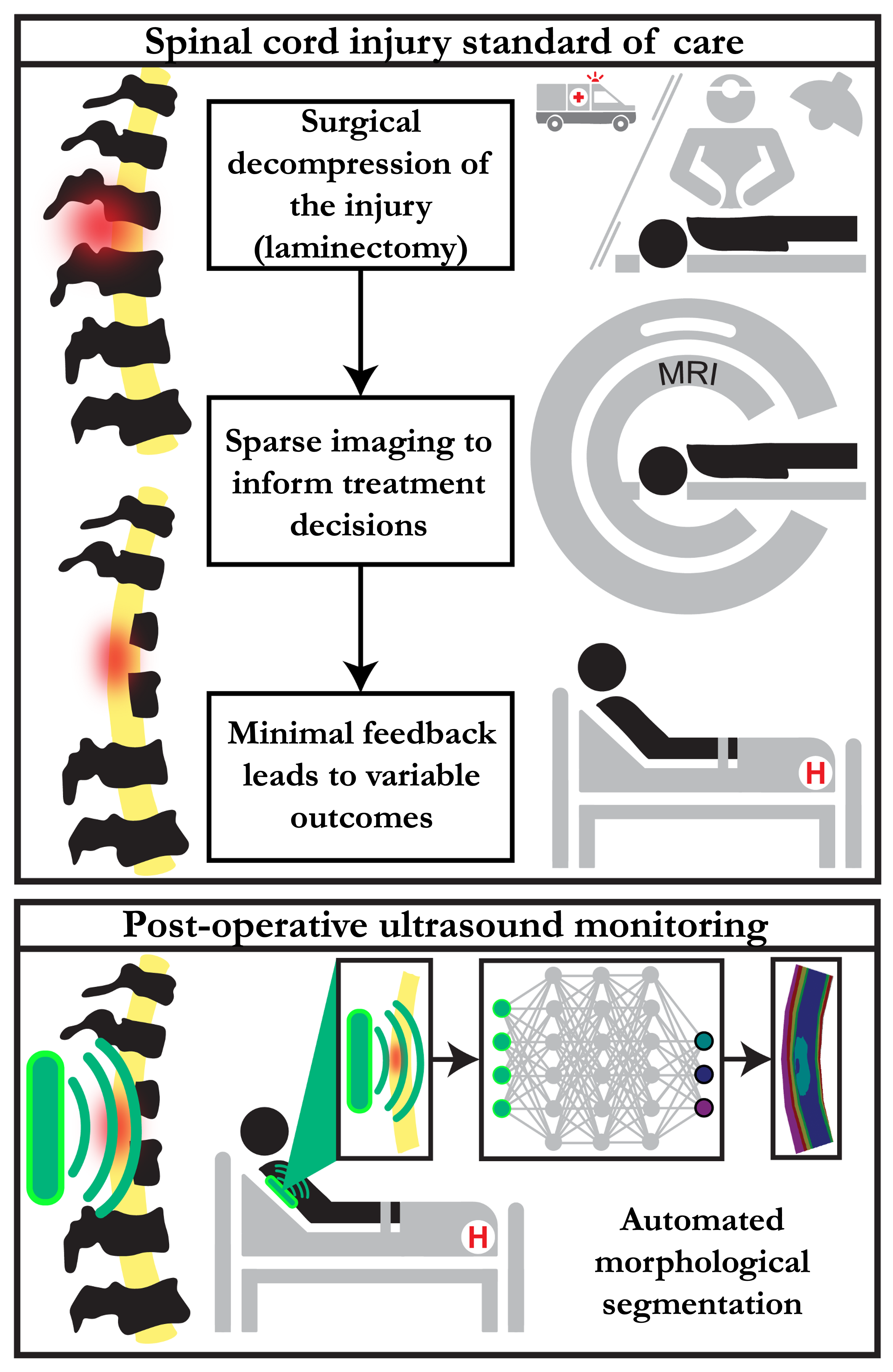}}
\caption{A comparison of the current standard of care for spinal cord injury with our approach using ultrasound imaging and deep learning for automatic diagnostics. Current treatment approaches do not provide a comprehensive and continuous avenue for monitoring patient spinal cord health after surgical decompression. With automatic injury localization and spinal cord segmentation, clinicians can capture changing biomarkers such as hematoma progression and tissue inflammation to personalize and better understand treatment approach.} 
\label{fig5}
\end{figure}

The algorithms were evaluated specifically for deployment on implantable or wearable ultrasound devices, which are ideal for long-term continuous diagnostics during secondary injury. Unlike the current clinical standard, where patient health monitoring through anatomical imaging is limited, our new dataset and benchmarking results enable a new paradigm for computer-aided diagnostics in SCI \cite{manbachi2020intraoperative}. With injury localization and anatomical segmentation, we can track injury progression and inflammation in tissue to understand patient health trajectory and optimally titrate interventional therapies (Figure \ref{fig5}). There are already several academic efforts on wearable and implantable ultrasound for clinical settings \cite{la2022flexible, hu2023stretchable, liu2024wearable}, and growing research interest in exploring innovative applications and methodologies at the intersection of wearable technologies and machine learning algorithms with over 40,000 papers published in 2023 with the keywords "wearable" and "ML" \cite{olyanasab2024leveraging}. Moreover, ML approaches can analyze long-term patterns from continuous data streams to uncover patterns that humans cannot due to the complexity and sheer volume of the data \cite{benke2018artificial, huang2023emerging}. These advantages can facilitate easier integration of ultrasound into robotic and tele-operative systems for remote diagnosis, presenting novel avenues for augmenting clinical insights and potentially improving the accessibility of medical care \cite{duan2021tele}. With this quantitative approach for image analysis, ML can also mitigate the issue of inter-observer variability. Furthermore, medical image segmentation of the spinal cord can aid in patient-specific computer simulations in the spinal cord to facilitate investigations in novel treatment paradigms using therapeutic focused ultrasound \cite{kumar2023patient, kumar2023computational}. 
\par  
It is important to note that there are some limitations to our study which we plan to improve in future works. Firstly, each segmentation model was pretrained on the ImageNet dataset, which contains millions of RGB images, before fine-tuning with our ultrasound dataset. We expect that fine-tuning with these grayscale medical images may cause some performance degradation in the resulting accuracy, which pretraining on a large medical dataset can improve. With this approach, the underlying learned features can have some foundational similarities with the dataset of interest. One such example of this is the Synapse multi-organ dataset, which contains CT images across several organs \cite{xu2016multi}. Checkpoints for specific models trained on medical images may also be applicable here \cite{kang2023deblurring}. Deep learning algorithms pretrained on ImageNet may learn to make decisions of boundaries between different segmentation regions based on the variance of intensity, which is reasonable in natural images, but less suitable for medical images. This is because medical images are primarily grayscale, which can result in lower intensity variance compared to the rich color information in natural images. The reliability and generalizability of deep learning models is also curtailed by the issue that images were generated by only two different types of transducers at a single collection site (T5), which can be improved with more diverse data collection processes. The accuracy of these models on the test may also increase with augmentation techniques specific for B-mode ultrasound images that use physics-inspired transformations, including deformation, reverb, and signal-to-noise ratio  \cite{tirindelli2021rethinking, al2019deep}. 
\begin{figure}
\centerline{\includegraphics[scale = 0.37]{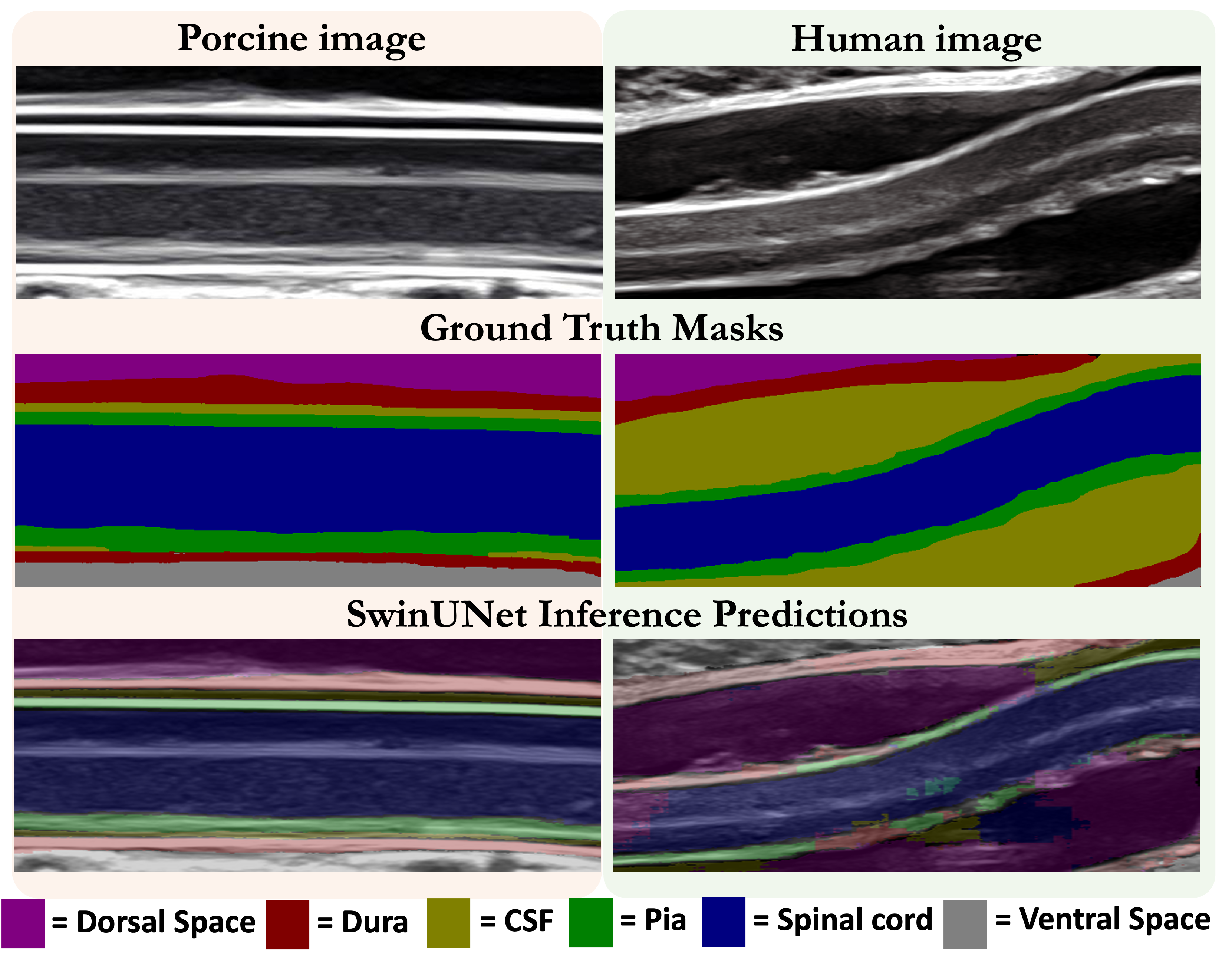}}
\caption{Visualization of SwinUNET's performance on an example porcine image and example human image.} 
\label{fig6}
\end{figure}
\par 
Additionally, the poor zero-shot generalization to human images of these models is influenced by the characteristics and processing of our collected human dataset. While human and porcine spine are morphologically similar, suggesting high generalization accuracy across these species, in the images collected, most patients were undergoing a spinal cord shortening procedure, significantly affecting the curvature of the cord. Even with these drastic changes to spinal cord geometry, SegFormer, TransUNet, SwinUNet, and SAMed were able to segment the spinal cord with high accuracy (Dice $>$ 0.74). However, because the CSF space is much larger in humans compared to porcine subjects (that were not undergoing PVCSO), after data preprocessing some anatomical classes were not visible in the frame, such as the dorsal space, dura, and ventral space (Figure \ref{fig6}). This may have a significant impact on the MIoU and Dice scores in Table \ref{tab:segmentation-results}, and should be further explored. Moreover, we had no examples of the hematoma, dura/pia compelex, or dura/ventral space classes in this human inference dataset. Future efforts to improve human generalization can include transfer learning techniques to adapt the deep learning model trained on porcine data for human applications. Employing domain adaptation techniques, such as matching component analysis, cannoncial correlation analysis, or optimal transport, can effectively augment porcine data into "synthetic" human data, enhancing the model's ability to generalize across species \cite{clum2020matching, andrew2013deep}. These approaches would rely on a large set of porcine images and a smaller set of human images that it can map onto, and the morphological similarities in these spinal cord anatomies are favorable as transfer learning typically performs best across domains with related underlying distributions \cite{clum2020matching, toossi2021comparative}. As ultrasound evaluation continues to become the standard-of-care in SCI, there will be increased datasets of human spinal cord images to better inform and train these algorithms \cite{aarabi2022proposal}. 

\section*{Conclusion}
The deployment of artificial intelligence and computer vision in ultrasound image analysis holds remarkable potential for streamlining diagnostics, with significant enhancements in image quality, diagnostic precision, and accessibility. However, effectively training deep learning models for this purpose presents various challenges, including the need for large, high-quality annotated datasets. We hope that publicly releasing this unique dataset will further facilitate computer vision efforts in medical imaging. By automating diagnostics within ultrasound, clinical workflows for personalized treatment paradigms can be augmented without overburdening clinicians. The significance of this dataset and research effort is highlighted by ultrasound's unique capability for real-time and noninvasive imaging, providing continuous insights on patient health. With the rapidly evolving field of ultrasound imaging, enabling high-resolution and dense datasets, the benefits of deep learning can be realized across diverse healthcare settings.

\bibliography{sample}

\section*{Acronyms}
\begin{longtable}{>{\raggedright\arraybackslash}p{5cm} >{\raggedright\arraybackslash}p{12cm}}
AR & Average Recall \\
ASPP & Atrous Spatial Pyramid Pooling \\
B-mode & brightness-mode \\
CNN & convolutional neural network \\
CPU & central processing unit \\
CSF & cerebrospinal fluid \\
CT & computed tomography \\
CVAT & Computer Vision Annotation Tool \\
DETR & Detection Transformer \\
DICOM & Digital Imaging and Communications in Medicine \\
E-ELAN & Extended Efficient Layer Aggregation Network \\
FPS  & frames per second \\
GPU & graphics processing unit \\
IoU & Intersection over Union \\
mAP & mean Average Precision \\
MIoU & Mean Intersection over Union \\
ML & machine learning \\
MRI & magnetic resonance imaging \\
PNG & Portable Networks Graphics \\
PVCSO & posterior vertebral column subtraction osteotomy \\
RCNN & Region-based Convolutional Neural Network \\
SAMed & Segment Anything for medical images \\
SCI & spinal cord injury \\
SSD & Single Shot Detector \\ 
YOLO & You Only Look Once
\end{longtable}

\section*{Acknowledgements}
This work received funding support from the National Science Foundation (NSF) STTR Phase 1 Award (\#:1938939), Defense Advanced Research Projects Agency (DARPA) Award (\#:N660012024075), National Institutes of Health (NIH) awards T32 GM136577 and F30 HL168823, and Johns Hopkins Institute for Clinical and Translational Research (ICTR)’s Clinical Research Scholars Program (KL2), administered by the NIH National Center for Advancing Translational Sciences (NCATS).

\section*{Author contributions statement}
AK - project conception, data processing and annotation, deep learning, manuscript preparation, figures. KK - data processing and annotation, deep learning. KJ, MB, DD, CWL, SK, JS, KA - data annotation. MK - data collection, rendered Fig. 6. RL, KKL - manuscript review. AH, DR - data collection. IS - rendered Fig. 1. BT, NT, NT, AM - funding acquisition, supervision. 

\section*{Additional Information}
\subsection*{Competing interests}
The authors declare no competing interests.

\subsection*{Data Availability}
The authors will maintain this dataset with changes and updates to be described on the following GitHub page: \url{https://github.com/HEPIUSLAB/ultrasound_spinal_cord_dataset}. 

\subsection*{Supplementary Materials}
The supplementary materials for this manuscript are available on the following page: \url{https://drive.google.com/file/d/1cY400awTul8FEVQAm9LMCUouuuzCGT3d/view?usp=sharing}

\end{document}